\documentclass[nanomaterials,article,accept,moreauthors]{mdpi} 
\graphicspath{{Definitions/}}

\firstpage{1} 
\pubvolume{9}
\issuenum{5}
\articlenumber{663}
\pubyear{2019}
\copyrightyear{2019}
\history{Received: 15 March 2019; Accepted: 19 April 2019; Published: 26 April 2019}
\Title{Thermal Boundary Characteristics of Homo-/Heterogeneous Interfaces}

\Author{Koen Heijmans $^{1}$\orcidA{}, Amar Deep Pathak $^{1}$\orcidB{}, Pablo Solano-L\'{o}pez $^{2}$, Domenico Giordano $^{3}$, Silvia~Nedea $^{1,}$* and David Smeulders $^{1}$}
\AuthorNames{Koen Heijmans, Amar Deep Pathak, Silvia Nedea, Pablo Solano-L\'{O}pez, Domenico Giordano, and David Smeulders}

\address{%
$^{1}$ \quad Energy Technology, Department of Mechanical Engineering, Eindhoven University of Technology, \mbox{ 5600 MB Eindhoven}, The Netherlands; k.heijmans@tue.nl (K.H.); simplyamar06@gmail.com (A.D.P.); d.m.j.smeulders@tue.nl (D.S.) \\
$^{2}$ \quad Departmento de Fisica Aplicada, ETSIAE, Universidad Polit\'{e}cnica de Madrid, 28040 Madrid, Spain; Pablo.Solano@upm.es\\ %Please add city and post code
$^{3}$ \quad ESA---Estec, Keplerlaan 1, 2201 AZ Noordwijk, The Netherlands; dg.esa.retired@gmail.com}
\corres{Correspondence: s.v.nedea@tue.nl}

\abstract{{The interface of two solids in contact introduces a thermal boundary resistance (TBR), which is challenging to measure from experiments. Besides, if the interface is reactive, it can form an intermediate recrystallized or amorphous region, and extra influencing phenomena are introduced. Reactive force field Molecular Dynamics (ReaxFF MD) is used to study these interfacial phenomena at the (non-)reactive interface. The non-reactive interfaces are compared using a phenomenological theory (PT), predicting the temperature discontinuity at the interface. By connecting ReaxFF MD and PT we confirm a continuous temperature profile for the homogeneous non-reactive interface and a temperature jump in case of the heterogeneous non-reactive interface. ReaxFF MD is further used to understand the effect of chemical activity of two solids in contact. The selected Si/SiO$_2$ materials showed that the TBR of the reacted interface is two times larger than the non-reactive, going from $1.65\times 10^{-9}$ to $3.38\times 10^{-9}$ m$^2$K/W. This is linked to the formation of an intermediate amorphous layer induced by heating, which remains stable when the system is cooled again. This provides the possibility to design multi-layered structures with a desired TBR.}}


\begin{document}
%%%%%%%%%%%%%%%%%%%%%%%%%%%%%%%%%%%%%%%%%%

%%%%%%%%%%%%%%%%%%%%%%%%%%%%%%%%%%%%%%%%%%
%\setcounter{section}{-1} %% Remove this when starting to work on the template.
%\section{How to Use this Template}
%The template details the sections that can be used in a manuscript. Note that the order and names of article sections may differ from the requirements of the journal (e.g., the positioning of the Materials and Methods section). Please check the instructions for authors page of the journal to verify the correct order and names. For any questions, please contact the editorial office of the journal or support@mdpi.com. For LaTeX related questions please contact latex@mdpi.com.
%The order of the section titles is: Introduction, Materials and Methods, Results, Discussion, Conclusions for these journals: aerospace,algorithms,antibodies,antioxidants,atmosphere,axioms,biomedicines,carbon,crystals,designs,diagnostics,environments,fermentation,fluids,forests,fractalfract,informatics,information,inventions,jfmk,jrfm,lubricants,neonatalscreening,neuroglia,particles,pharmaceutics,polymers,processes,technologies,viruses,vision

\section{Introduction}
\begin{figure}[H]
	\centering
	\includegraphics{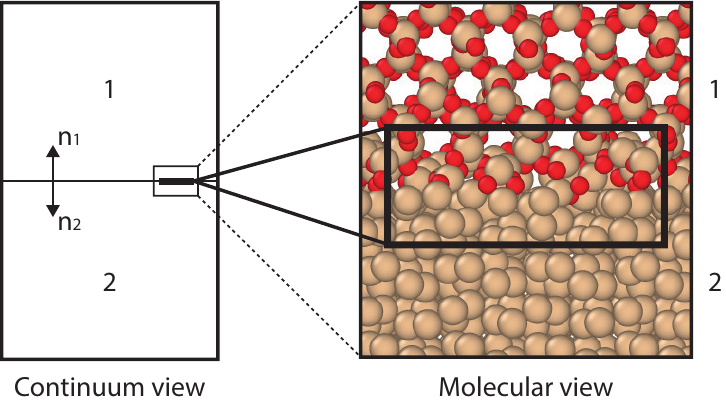}
	\caption{\label{fig:models}A schematic representation of continuum and molecular view of a system chosen for the present~study.}
\end{figure}

Molecular characteristics of solids in contact play a key role in various fundamental studies related to heat transfer~\cite{samvedi2009role}, mechanical behavior~\cite{pilania2014revisiting}, micro/nano-fluidics~\cite{kim2015molecular}, and~catalysis~\cite{mueller2010development}. Besides, it plays an important role in applications related to semiconductors~\cite{majumdar1993microscale}, microelectronics~\cite{schutze2001integrated,nunzi2008adsorption} and heat-shielding in re-entry vehicles for aerospace applications~\cite{giordano2017exploratory}. In~the latter case, insight into the thermal resistance at an interface of a multilayer structure is crucial for prediction and control of overheating of the thermal protection system used for the re-entry shuttle. The~non-equilibrium effects of a hypersonic flow impinging on a solid interface requires detailed investigation of the boundary processes. Mass, momentum, energy transfer and chemical reactions on the interface are critical under these extreme conditions~\cite{dias2016development}. These processes can change the interfacial properties significantly compared to initial bulk properties. Furthermore, in~case of chemical reactions at the interface (as shown in Figure~\ref{fig:models}), insight into the heat transfer in a small layer of material (a few molecular layers) is~required.\par
The influence of the solid interfaces on thermal properties can be analyzed by the local temperature profiles on a molecular level. Experimental measurement of a temperature profile at the molecular level is extremely challenging. Therefore, computational models can be useful accurate tools to provide thermal insight and to establish the interfacial thermal correlations. In~the context of building up a macroscopic theory of gas--surface interactions targeting the hypersonic re-entry flows, Giordano~et~al.~\cite{giordano2017exploratory} have proposed a Phenomenological-Theory (PT) to study heat transfer between two solids in contact. Another tool is Molecular Dynamics (MD) simulations, which has been used before to investigate thermal transport across solid interfaces~\cite{merabia2012thermal,mahajan2011estimating,deng2014kapitza,lampin2012thermal,chen2012thermal,zushi2015effect,schelling2002comparison}, and~significant influences of the solid interface on the thermal conduction are reported. A~schematic representation of these two methods is shown in Figure~\ref{fig:models}, with~a continuum view for PT, and~a molecular view for~ReaxFF. 

Though temperature profiles at the solid interface are investigated before, these studies and methods focus mainly on non-reactive interfaces. However, chemical activity at the interfaces can influence the thermal behavior of solids in contact. Therefore, reactive force field Molecular Dynamics (ReaxFF~\cite{ReaxFFAdriFirst,senftle2016reaxff}) is used, which is able to capture the chemical reaction and its influence on the surface transformation and temperature profile on molecular~level.

In this study, we first consider the characterization of material properties like thermal expansion, thermal conductivity, and~elastic properties using ReaxFF, to~validate the force field. Thereafter, we build a generic non-reactive system, in~which an interface is created based on the same material, for~this, we considered two Platinum slabs (homogeneous Pt/Pt interface). Further, we create an interface between two different materials in contact, a~non-reactive heterogeneous Platinum-Nickel interface (Pt/Ni).  Accordingly, the~temperature profiles from the ReaxFF MD simulations are compared with the macroscopic level PT based model of Giordano~et~al.~\cite{giordano2017exploratory}. For~the computation of the temperature profile with PT, relevant material properties are computed with ReaxFF MD and upscaled, to~be used as input in the PT model. Platinum and Nickel were selected because they are well studied, non-reactive, monatomic, and~have similar lattice~size. %when we talk about the interface between two materials, the - is replaced by a /

After we analyzed this generic model, we created the reactive heterogeneous interface of Silicon and Silicon-oxide (Si/SO$_2$). This Si/SiO$_2$ interface is of relevance for many applications in the semiconductor industry, as~well as aerospace engineering~\cite{kulkarni2012oxygen}.
{Because of its relevance, the~Si/SiO$_2$ interface is studied numerous times before~\cite{van2003reaxffsio}, including the thermal boundary resistance (TBR) based on experiments~\cite{hurley2011measurement}, or~numerical methods like MD~\cite{deng2014kapitza,lampin2012thermal,chen2012thermal}, acoustic and diffusive mismatch models (AMM, DMM)~\cite{hu2001thermal}, and~phonon wave-package method~\cite{deng2014kapitza}. With~ranging values between 0.4--3.5 $\times 10^{-9}$ m$^2$K/W, depending on the method and composition of the materials. Furthermore, Chen~et~al. showed a strong correlation between the coupling between the materials and the TBR. These coupling can be directly related to reactions happening at the interface. However, to~our knowledge no systematic studies has been done on the influence of a reactive interface on the TBR. Therefore, we used ReaxFF to study the influence of the reactive interface. The~Si/SiO$_2$ system is kept at various temperatures, within~the ReaxFF simulations, to~increase/decrease the chemical activity. Accordingly, the~TBR is computed. The~TBR is defined as the temperature discontinuity at the interface ($\Delta T$) divided by the heat flux ($Q$) that crosses the interface, see Equation~(\ref{TBR}). The~TBR is often referred as the Kapitza resistance~\cite{kapitza1941heat}, however, we kept the analogy of Peterson~et~al. \cite{peterson1973kapitza} }
\begin{equation}\label{TBR}
TBR = R = \Delta T/Q
\end{equation}

\section{Methodology \& Material Properties~Estimation}\label{sec_methodology}
\vspace{-6pt}
\subsection{Phenomenological-Theory (PT)}
To study the heat transfer between two solids in contact, Giordano~et~al.~\cite{giordano2017exploratory} have proposed a Phenomenological-Theory (PT). With~the aim of building up a macroscopic theory of gas--surface interactions targeting the hypersonic re-entry flows. They have remarked the lack of a physical principle justifying the standard temperature-continuity boundary conditions as a replacement of temperature-continuity, and~have introduced tension continuity:
\begin{equation}
\textbf{n}_1\cdot \tau_{U,1}(P,t) + \textbf{n}_2\cdot \tau_{U,2}(P,t) = 0
\label{mfc}
\end{equation}
where, $\textbf{n$_1$}$ and $\textbf{n$_2$}$ are normal unit vectors at a point of contact $P$ and time $t$. This macroscopic theory is founded on momentum conservation and represents a more physically motivated boundary condition. For~the mathematical formulation of the phenomenological-theory and further details, readers can refer to the original paper~\cite{giordano2017exploratory}. This method is used to compare the temperature profile non-reactive interfaces studied with MD. For~the input of required material properties, the~MD calculated values are~used.

\subsection{Reactive Force Field Molecular~Dynamics}

Molecular Dynamics (MD) is a computational method to obtain macroscopic and microscopic properties from approximated trajectories of individual particles. These approximated trajectories, obtained from Newton's equations of motion, form an ensemble from which macroscopic properties of materials can be obtained~\cite{frenkel2001understanding}. To~capture the chemical change during a reaction, Reactive force field (ReaxFF)~\cite{ReaxFFAdriFirst} is used. ReaxFF is computationally more expensive than the non-reactive force field, however, it allows bond formation and bond breaking during the simulations, which makes simulations of chemical reactions possible. According to ReaxFF the bond order between a pair of atoms can be obtained directly from the inter-atomic distance, which relation is used to mimic chemical change. {The feature of bond formation and breaking allows the user not to give predefined reactions pathways, these should present themselves given the right temperatures and chemical environment. However, the~accuracy of this relies directly on the training set and the weights that are used to parameterize the reactive force field. Therefore, we tested several available ReaxFF on their ability to predict relevant material characteristics for our study.} ReaxFF is widely used in studying chemical activities at a molecular level~\cite{senftle2016reaxff,pathak2016reactive}, including many Si/SiO$_2$ systems~\cite{kim2014development,suek2018characterization,nielson2005development,kulkarni2012oxygen,fogarty2010reactive,narayanan2011reactive,castro2013comparison,pitman2012dynamics,zou2012investigation,newsome2012oxidation,van2003reaxffsio}.

For the {in-silico} characterization using ReaxFF MD methodology, we compute the material properties of relatively simple systems of Platinum (Pt), Nickel (Ni), Silicon (Si) and Silicon dioxide (SiO$_2$). Furthermore, the~elastic properties, thermal expansion coefficient, and~thermal conductivity of the materials are important parameters for the phenomenological-theory, and~thus the computed values are used as input parameters to upscale the molecular results up to the macroscopic~level.

\subsubsection{Force Field~Selection}
Selection of an appropriate force field is very important for {in-silico} characterization. The~calculations of elastic properties, thermal expansion coefficients and the radial distribution function (RDF), guide as a selection criterion for the appropriate force fields. Supercells of $(5\times 5\times 5)$ Pt, $(5\times 5\times 5)$ Ni, $(3\times 3\times 3)$ Si and $(4\times 4\times 4)$ SiO$_2$ are created, containing approximately 400--800 atoms, with~initial volumes of 7547, 5469, 7890, and~10,913 \AA$^3$, respectively. Periodic Boundary Conditions (PBC) are applied in all directions. The~unit cells of Pt~\cite{wyckoff1963wg}, Ni~\cite{swanson1953standard}, Si~\cite{kasper1964crystal} and SiO$_2$~\cite{nieuwenkamp1935kristallstruktur} are taken from experimental crystallographic information~files. \par
\begin{figure}[H]
	\centering
	\includegraphics{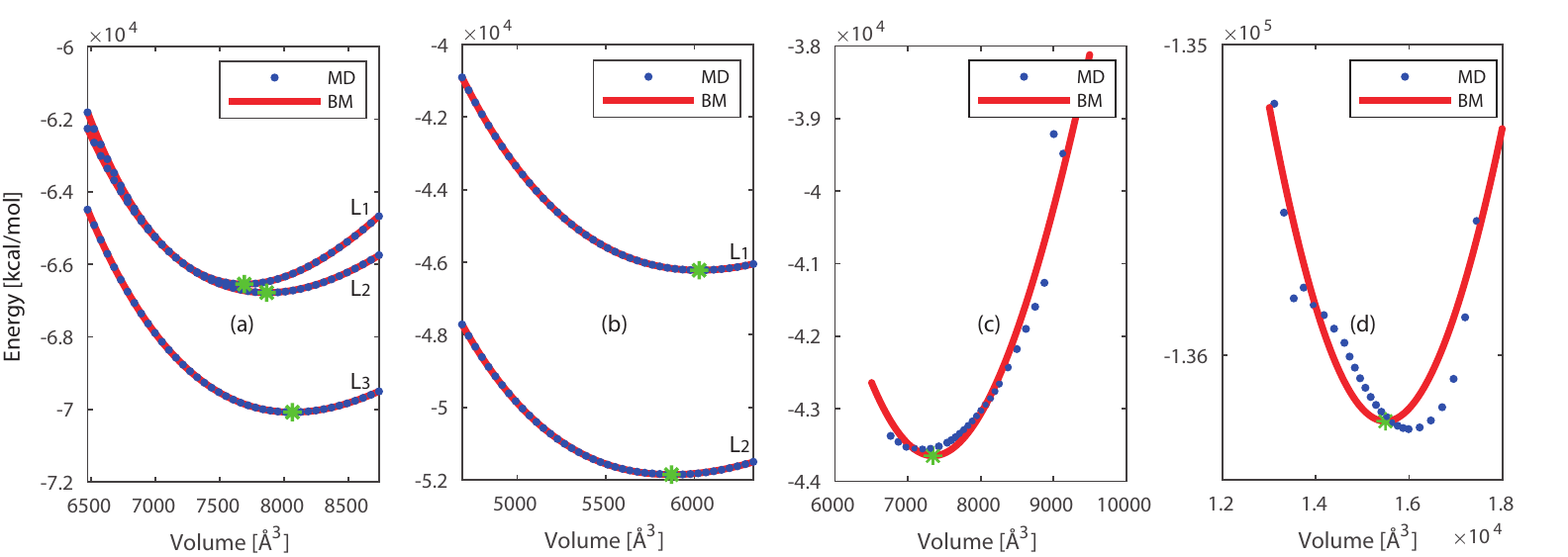}
	\caption{\label{fig:BMeos}The energy of system obtained from selected force fields as a function of volume and the fitted BM-eos for: (\textbf{a}) $5 \times 5 \times 5$ Pt, (\textbf{b}) $5 \times 5 \times 5$ Ni, (\textbf{c}) $3 \times 3 \times 3$ Si, and~(\textbf{d}) $4 \times 4 \times 4$ SiO$_2$. The~\lq$\star$\rq represents the minimum energy point ($E_0, V_0$) on each~curve.}
\end{figure}

  
We have chosen three reactive force fields~\cite{mueller2010development,sanz2008molecular,nielson2005development} available for Pt and Ni and nine force fields for Si and SiO$_2$~\cite{nielson2005development,kulkarni2012oxygen,fogarty2010reactive,narayanan2011reactive,castro2013comparison,pitman2012dynamics,zou2012investigation,newsome2012oxidation}. These force fields are tested by deforming the crystals in the range of 0.86 to 1.16 times their initial volume. The~resulting increase in potential energy is shown in Figure~\ref{fig:BMeos}.
%\begin{table}
%	\centering
%	\caption{\label{tab:BM-eos}Comparison of computed bulk modulus and relaxed volume ($V_0$) from various reactive force fields, the corresponding values from literature are given in parentheses.}
%	%\begin{ruledtabular}
%		\begin{tabular}{cccccc}
%			Material & Force field &Fig.&$E_0$&$B_0$& $V_0$ \\ 
%			& reference&&[kcal/mol/unit cell]&[GPa]&[\AA$^3$]\\ \hline
%			Pt& ~\cite{mueller2010development} &\ref{fig:BMeos}a L1&-532.4&240 (266~\cite{holmes1989equation})&61.52 (60.38~\cite{wyckoff1963wg}) \\ 
%			Pt& ~\cite{sanz2008molecular}&\ref{fig:BMeos}a L2&-534.3&179 (266~\cite{holmes1989equation})&62.91 (60.38~\cite{wyckoff1963wg})\\ 
%			Pt& ~\cite{nielson2005development} &\ref{fig:BMeos}a L3&-560.6&166  (266~\cite{holmes1989equation})&64.52 (60.38~\cite{wyckoff1963wg})\\ 
%			Ni& ~\cite{mueller2010development} &\ref{fig:BMeos}b L2&-414.9&155 (185~\cite{chen2000compressibility})&46.96 (43.76~\cite{swanson1953standard})\\ 
%			Ni& ~\cite{nielson2005development} &\ref{fig:BMeos}b L1&-369.8&167 (185~\cite{chen2000compressibility})&48.21 (43.76~\cite{swanson1953standard}) \\ 
%			Si& ~\cite{kulkarni2012oxygen} &\ref{fig:BMeos}c&-1617&144 (117~\cite{PhysRevBSiDFT})&272.1 (292.2~\cite{kasper1964crystal}) \\ 
%			SiO$_2$& ~\cite{kulkarni2012oxygen} &\ref{fig:BMeos}d&-2128&35 (36~\cite{liu1993bulk})&242.1 (170.5~\cite{nieuwenkamp1935kristallstruktur})\\  
%			\hline  
%		\end{tabular}
%	%\end{ruledtabular}
%\end{table}
For clarity only the best performing force field is shown for Si and SiO$_2$. The~relation between volume and energy is found by integration of the pressure in the third order Birch-Murnaghan equation of state (BM-eos)~\cite{murnaghan1944compressibility,birch1978finite,colonna2011high}, this relation is fitted to the results of the deformed crystal. The~resulting parameters ($B_0$ and $V_0$) are given in Appendix~\ref{app:BM-eos} and compared with the~literature. 

{The force field developed by J.E. Mueller~et~al.~\cite{mueller2010development} showed the best results, and~was therefore chosen for further use including Pt and Ni. This force field was parameterized for studying hydrocarbon reactions on nickel surfaces. They included the equation of state (EOS) for different Ni bulk structures in the training. Both our and their calculations predicted the EOS, together with the lattice parameters, in~close agreement with quantum mechanical calculations. We predicted the equilibrium volume ($V_0$) of Pt~\cite{wyckoff1963wg} and Ni~\cite{swanson1953standard} unit cells within 1.9\% and 7.0\% deviations from their experimentally observed crystal structures. The~deviations for the bulk modulus ($B_0$) of Pt and Ni are 9.8\%~\cite{holmes1989equation} and 16.0\%~\cite{chen2000compressibility} respectively. Furthermore, Mueller~et~al. computed cohesive energies in close agreement with experimental values. The~force field developed by Kulkarni~et~al.~\cite{kulkarni2012oxygen} was chosen for further use including  Si and SiO$_2$. This force field is an extension to include gas--surface reactions between oxygen and silica into an existed force field developed by van Duin~et~al.~\cite{van2003reaxffsio}. This original force field was parametrized to include the chemistry of silicon and silicon oxides, and~the interface between these materials. Previous work of Tian~et~al.~\cite{tian2017thermal} also indicated that this original force field predicts the thermal conductivity of vitreous SiO$_2$ in close agreement with experimental values. The~force field of Kulkarni~et~al. is able to predict the equilibrium volume ($V_0$) within 7.5\%~\cite{kasper1964crystal} and 30\%~\cite{nieuwenkamp1935kristallstruktur} deviations for Si and SiO$_2$, respectively. The~bulk modulus ($B_0$) is within 22.9\%~\cite{PhysRevBSiDFT} and 5.7\%~\cite{liu1993bulk} deviations for Si and SiO$_2$, respectively. Therefore, this force field is selected for the study that includes Si and SiO$_2$. These force fields are chosen for further investigation.}
%please define

\begin{figure}[H]
	\centering
	\includegraphics{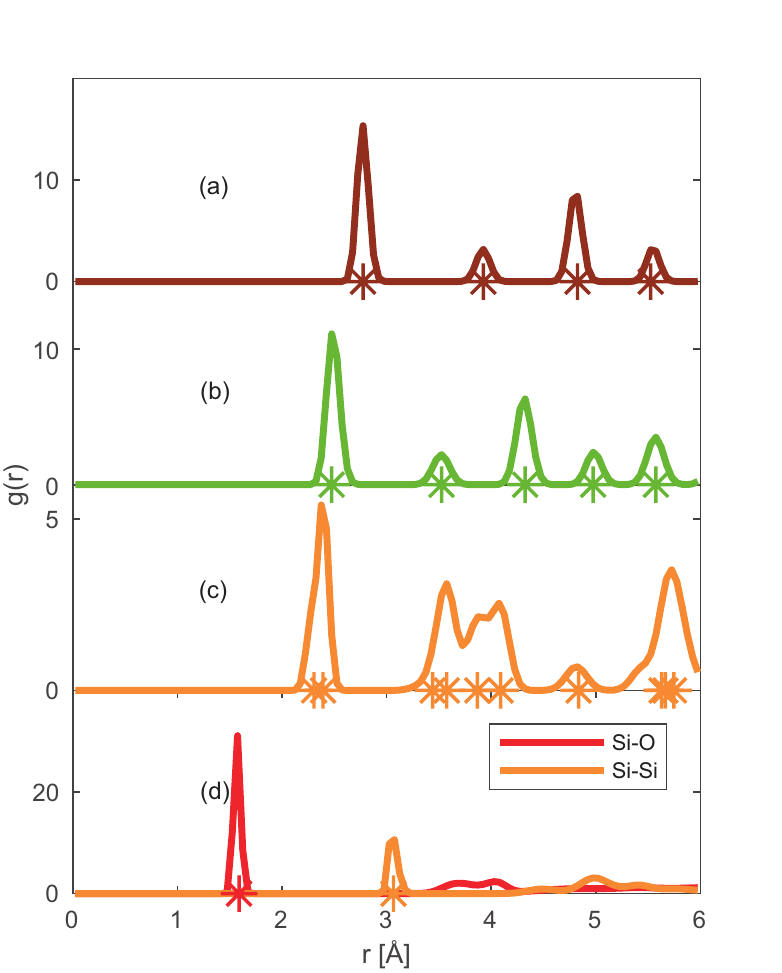}
	\caption{\label{fig:RDF}Radial distribution functions for: (\textbf{a}) Pt--Pt, (\textbf{b}) Ni--Ni, (\textbf{c}) Si--Si and (\textbf{d}) Si--O pairs present in studied solid crystals obtained from the ReaxFF MD simulations. The~\lq$\ast$\rq~represents the neighboring atomic distances inside the solid crystal. For~clarity, only the nearest neighbor \lq$\ast$\rq~is shown in SiO$_2$.}
\end{figure}

To validate further the applicability of the chosen force fields, we have obtained the Radial Distribution Functions (RDFs) of Pt--Pt, Ni--Ni, Si--Si (Si and SiO$_2$) and Si--O (SiO$_2$) pairs from ReaxFF MD simulations in periodic solid supercells as shown in Figure~\ref{fig:RDF}. 
The sharp peaks in RDF elucidate the extent of ordering in the supercell, thus representing the solid phase. The~locations of the peaks coincide with the position of neighboring atoms (represented by \lq$\ast$\rq) in the experimentally observed solid crystal~\cite{wyckoff1963wg,swanson1953standard,kasper1964crystal,nieuwenkamp1935kristallstruktur}. Concluding that the selected force fields~\cite{mueller2010development,kulkarni2012oxygen} can capture the crystalline phase of Pt, Ni, Si, and~SiO$_2$.\par
The volumetric thermal expansion coefficient ($\alpha_v$) can be obtained from the slope of the natural logarithm of the volume ($\ln V$) versus imposed temperature ($T$) \cite{mashreghi2012determining}:
\begin{equation}\label{eq:thermal_exp}
\alpha_v=\dfrac{1}{V}\left(\dfrac{\partial V}{\partial T}\right)_p = \Bigg[\dfrac{\partial ln(V)}{\partial T}\Bigg]_p
\end{equation}
where $\alpha_v$ is the volumetric thermal expansion coefficient at constant pressure. We have varied the temperature over 250--500 K at atmospheric pressure in an NPT ensemble. The~thermal expansion coefficients for Pt and Ni computed from ReaxFF MD and existing literature values are given in Table~\ref{tab:lin_exp_coeff}. The~results on the molecular scale are in reasonable agreement with the bulk experimental value~\cite{bolz1973crc,kollie1977measurement}.
\begin{table}[H]
	\centering
	\caption{\label{tab:lin_exp_coeff}Comparison of volumetric expansion coefficients ($ \alpha_v$) obtained from the present ReaxFF MD simulations, and~reported from literature (in parentheses).}
	%\begin{ruledtabular}
		\begin{tabular}{ccc}
		\toprule
			\textbf{Element}  & \textbf{\boldmath{$\alpha_v \times 10^{-5}$ (m$^3$/m$^3$K)}}& \\
			\midrule  
			Pt & 2.0 (2.7~\cite{bolz1973crc})&\\
			Ni & 3.2 (3.9~\cite{kollie1977measurement})&\\
			\bottomrule  
		\end{tabular}
%	\end{ruledtabular}
\end{table}
\unskip

\subsubsection{Thermal Conductivity with Steady State NEMD~Method}\label{sec:NEMD}
\begin{figure}[H]
	\centering
	\includegraphics{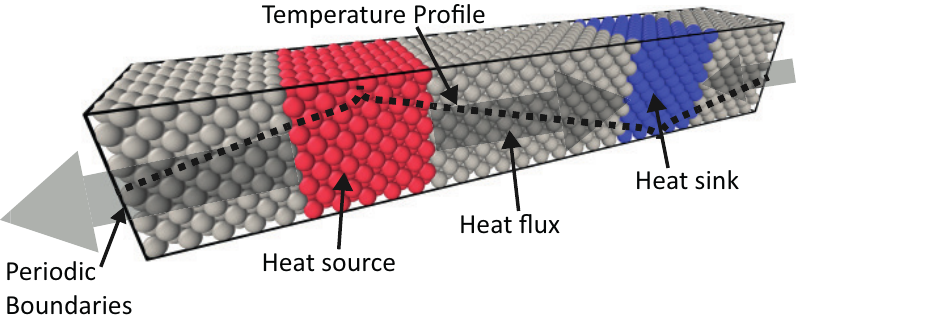}
	\caption{\label{fig:ss-NEMD}Schematic representation of the chosen system for steady state NEMD simulations. The~heat source and sink are coupled to different temperatures with a damping constant of 100 fs. This will induce a heat flux through the intermediate zone, which is weakly coupled with a damping constant of 100.000 fs. }
\end{figure}

The thermal conductivity of a solid can be computed from Steady State Non-Equilibrium Molecular Dynamics (ss-NEMD). This method has been previously used~\cite{chantrenne2004finite,schelling2002comparison,stackhouse2010theoretical,zhou2009towards,heino2003thermal,tian2017thermal}, and~it is based on imposing a temperature gradient over a system to estimate the thermal conductivity. A~schematic view of this method is shown in Figure~\ref{fig:ss-NEMD}. 
Two strongly coupled regions (using a Berendsen thermostat with damping constant $\tau$ = 100 fs) are created, one hot zone (red zone, $T_H$ = 330 K) and one cold zone (blue zone, $T_C$ = 300 K), which act as the heat source and sink, respectively. In~between these two zones, there are weakly coupled regions (gray zone, $\tau$ = 10$^5$ fs), this damping constant proved to have a negligible low influence on the system. This results in a steady state temperature gradient ($dT/dx$) and an energy flux ($q$), in~the weakly coupled regions between heat source and~sink.\par
From the energy flux and the temperature gradient, the~thermal conductivity ($k$) can be computed, following Fourier's law. This intuitive principle makes NEMD well suited to study thermal conductivity of different matter, and~investigate the influence of structural defects and solid interfaces~\cite{chantrenne2004finite}.
According to the kinetic theory, the~thermal conductivity ($k$) is related to the mean free path ($\lambda$) of energy carriers (Equation~(\ref{eq_Kcorrected})). If~the characteristic length of the system is larger than the mean free path of carriers, thermal energy is transferred by multiple collisions. In~this diffusive regime, the~Fourier law is still valid. In~cases when the characteristic length of the system is in the order of the mean free path, the~energy carriers may travel ballistically between source and sink. This scattering in the heat source and sink introduces an extra limiting effect on the mean free path, and~thus a reducing effect on the conductivity (Equation~(\ref{eq_Kcorrected})). Thus, the~conductivity equation must be corrected for the enhanced scattering effect~\cite{chen2005nanoscale}:
\begin{equation}\label{eq_Kcorrected}
k = \frac{1}{3}C_vv\lambda_L
\end{equation}
where, $C_v$ is the heat capacity, $v$ the energy carrier velocity, and~$\lambda _L$ the corrected mean free path for a system of size $L$. This can be estimated from Matthiessen's rule~\cite{dames2004theoretical}, which states that the corrected resistivity is the sum of the intrinsic scattering and the scattering due to impurity. Thus, the~corrected mean free path can be expressed as a combined effect of the mean free path of bulk ($\lambda_{\infty}$) and length of the system ($L$) as:
\begin{equation}\label{eq_lambdacorrected}
\frac{1}{\lambda_L} = \frac{1}{\lambda_{\infty}} + \frac{4}{L}
\end{equation}

In a system with periodic boundary conditions, the~average distance for an energy carrier to scatter with the heat source or sink is $L/4$~\cite{schelling2002comparison}. Combining Equations~(\ref{eq_Kcorrected}) and~(\ref{eq_lambdacorrected}), the~thermal conductivity ($k_L$) of system size $L$ can be expressed as:
\begin{equation}\label{eq_KcorrectedFinal}
\frac{1}{k_L} = \frac{12}{C_vv}\frac{1}{L} + \frac{1}{k_{\infty}}
\end{equation}

In this equation, $\frac{1}{k_{\infty}}$ is the thermal conductivity of the bulk material. Bulk thermal conductivity ($k_{\infty}$) can be estimated by extrapolating the effective thermal conductivity obtained for small system sizes ($k_L$).

The thermal conductivities of Pt and Ni are computed using the ss-NEMD method for different system sizes. The~total length of the systems varied from $3 \times 3\times X$ (with $X$ from 24 to 156), including $3 \times 3\times 6$ zones as the heat source and sink. The~energy flux is taken from the average of the heat flux added and extracted by the two strongly coupled regions. The~temperature gradient is computed over the weakly coupled region, and~the system is equilibrated up to 1 ns with time steps ($\Delta t$) of 0.25 fs. The~by ss-NEMD computed thermal conductivity values are presented in Table~\ref{tab:Pt_con}, and~increases with system size for both Pt and Ni~systems. \par 

\begin{table}[H]
	\centering
	\caption{\label{tab:Pt_con}Thermal conductivities for different system sizes of Pt and~Ni.}
		\begin{tabular}{ccc}
		\toprule
			\multirow{2}{*}{\textbf{System}}  & \textbf{Conductivity} & \textbf{Conductivity} \\
			&  \textbf{Pt (W/mK)}   &  \textbf{Ni (W/mK)} \\
			\midrule  3$\times$3$\times$24  & $4.0 \pm 0.4$   & $7.7\pm 0.5$ \\ 
			3$\times$3$\times$36  & $5.9\pm 0.3$  & --- \\ 
			3$\times$3$\times$60  & ---     & $17.2\pm 0.6$ \\ 
			3$\times$3$\times$84  & $10.4\pm 0.3$   & $21.7\pm 0.6$ \\  
			3$\times$3$\times$108 & $15.0\pm 0.5$  & $24.4\pm 0.5$ \\
			3$\times$3$\times$132 & $16.3\pm 0.3$   & --- \\ 
			3$\times$3$\times$156 & $19.2\pm 0.5$   & --- \\ \midrule
			Extrapolated & $49.8\pm 10.5$ & $74.4\pm 9.2$ \\
			Literature & 71.6~\cite{CRC} & 90.7~\cite{CRC}\\ \bottomrule  
		\end{tabular} 
\end{table}

\begin{figure}[H]
	\centering
	\includegraphics[scale=0.95]{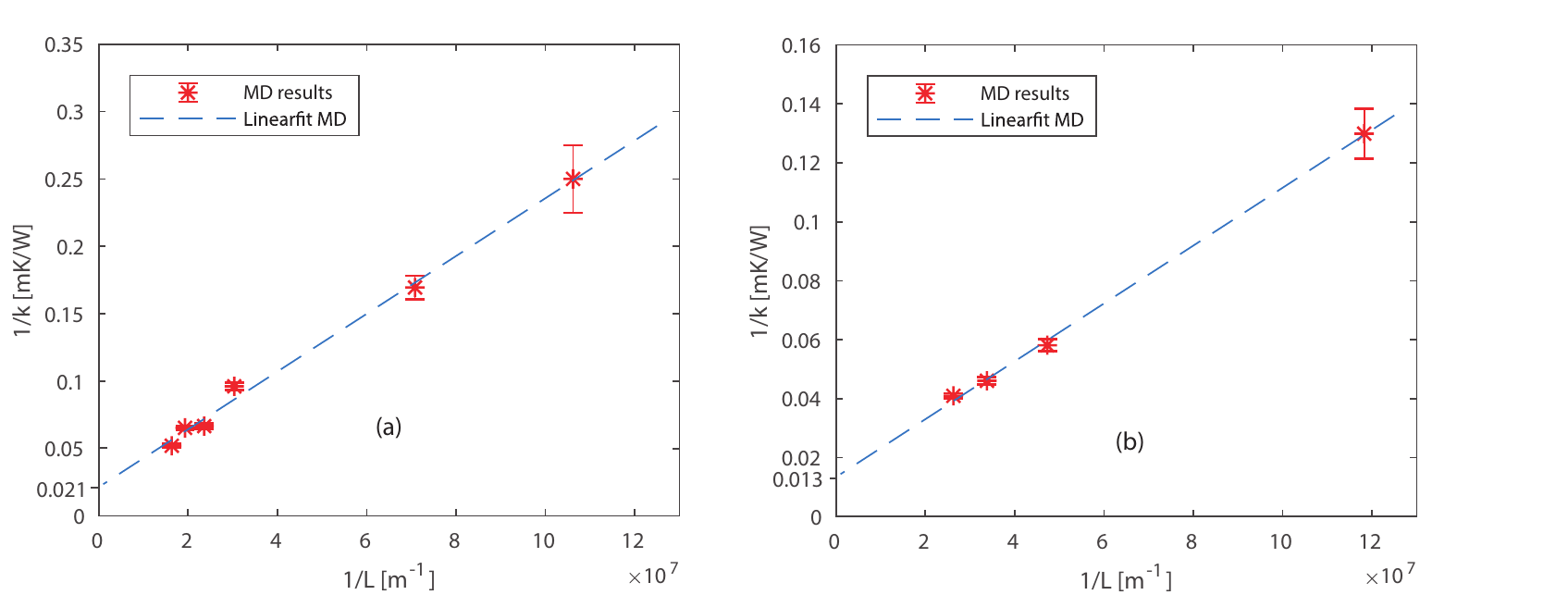}
	\caption{\label{fig:Extrapolation}Extrapolation of thermal conductivities for different sizes of (\textbf{a}) Pt and (\textbf{b}) Ni.}
\end{figure}

The thermal conductivities are extrapolated for long length by fitting the linear expression between $1/L$ and $1/k$ (Equation~(\ref{eq_KcorrectedFinal})) as shown in Figure~\ref{fig:Extrapolation}. The~fitted lines intersect the Y axis at $1/k_{\text{Pt}}$ = 0.021 mK/W and $1/k_{\text{Ni}}$ = 0.013 mK/W, which results in a bulk thermal conductivity of $k_{\text{Pt}}$ = 49.8 $\pm$ 10.5~W/mK and $k_{\text{Ni}}$ = 74.4 $\pm$ 9.2~W/mK. The~small deviations in an individual system result in large deviations in the bulk thermal conductivity, due to the extrapolation~\cite{berendsen2011student}. The~computed thermal conductivities are approximately 35 \% and 18 \% lower than literature~\cite{CRC} values for Pt and Ni, respectively. This was expected because the ReaxFF formalism we used does not describe free electrons. The~difference with experimental values can be explained by limitations of the ReaxFF method we used. ReaxFF MD is not able to model free electrons, thereby we underestimate the thermal conductivity. However, our aim is to compare models and confirm the temperature jump, not to compute exactly the thermal conductivities of Pt and Ni. There is a ReaxFF expansion including free electrons (e-ReaxFF) developed by Islam~et~al.~\cite{islam2016ereaxff}. At~the moment of writing this e-ReaxFF concept does not include the studied materials, nonetheless, this might be interesting for future~research.

The gradients of the extrapolated curve (Figure~\ref{fig:Extrapolation}) obtained from ReaxFF MD simulations are \mbox{2.1 $\times$ 10$^{-9}$ m$^2$K/W} and~1.0~$\times$ 10$^{-9}$ m$^2$K/W for Pt and Ni, respectively. The~gradient should be equal to the theoretical gradient $12/C_vv$ (Equation~(\ref{eq_KcorrectedFinal})) obtained from the kinetic theory. By~assuming a specific heat of $C_{v,\text{Pt}}$ = 29 $\times$~10$^{5}$~J/m$^3$K, and~$C_{v,\text{Ni}}$ = 40~$\times$~10$^{5}$~J/m$^3$K~\cite{CRC}, the~computed velocities of thermal transport carriers are $v_{\text{Pt}}\approx 2~\times10^{3}$~m/s, $v_{\text{Ni}}\approx3~\times~10^{3}$~m/s. These values are found to be in agreement with the speed of sound in the lateral direction through Pt and Ni from literature~\cite{CRC}. We~also studied the final size effects perpendicular to the heat flow for platinum systems, see~Appendix~\ref{app:size-effects_perp}. However, no finite size effects were observed for perpendicular directions, corresponding to the findings of Zhou~et~al.~\cite{zhou2009towards}.
\subsubsection{Building the Interfacial Molecular~System}
\begin{figure}[H]
	\centering
	\includegraphics[scale=0.95]{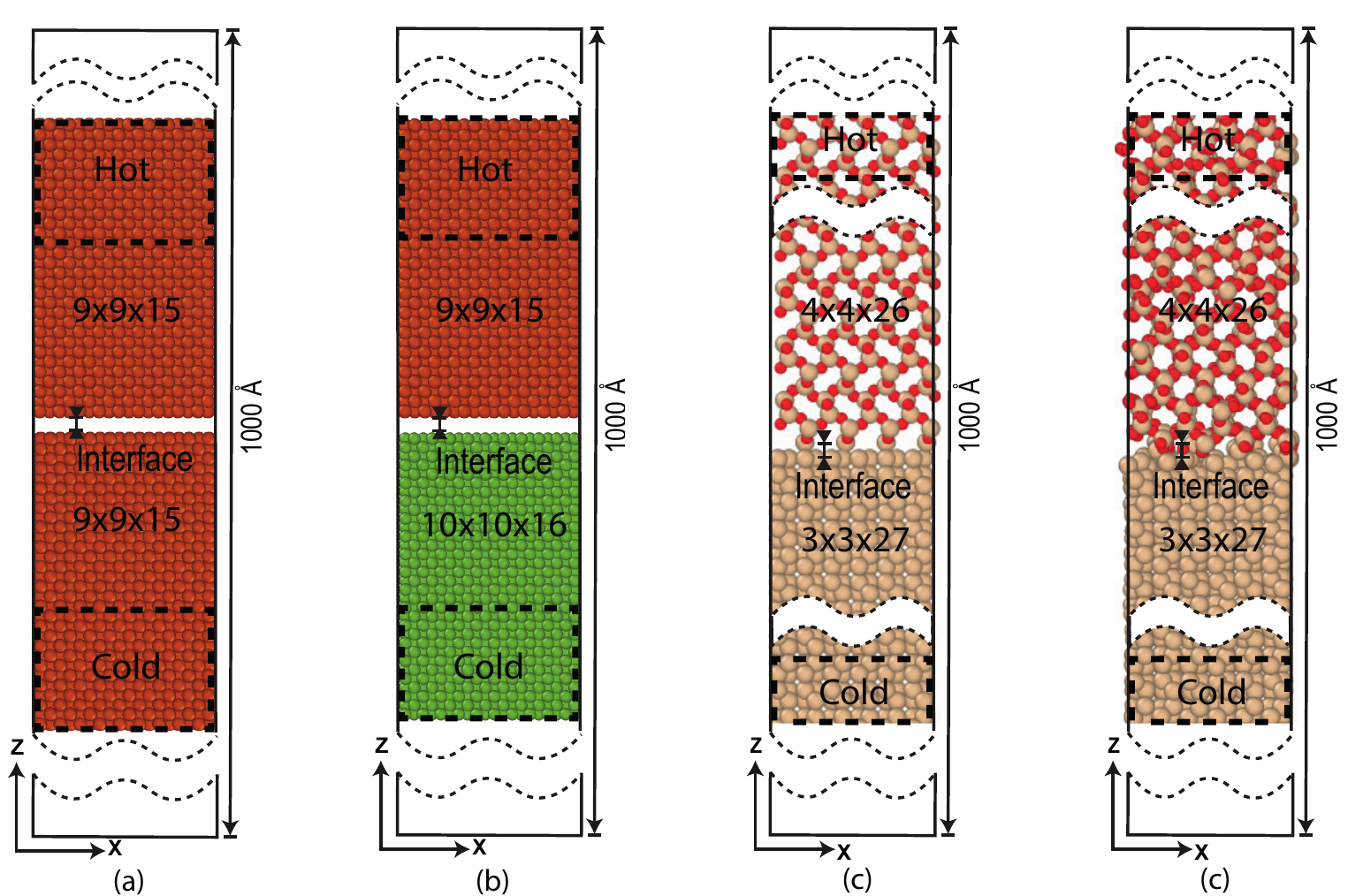}
	\caption{\label{fig:interfaces}Schematic representation of systems: (\textbf{a}) non-reactive Pt/Pt interface, (\textbf{b}) non-reactive Pt/Ni interface, (\textbf{c}) initial reactive Si/SiO$_2$ interface, and~(\textbf{d}) the merged reactive Si/SiO$_2$ interface. The~particles in the dashed area are the strongly coupled sections, which acts as a heat source and heat~sink.}
\end{figure}

Schematic diagrams of the studied Pt/Pt, Pt/Ni and Si/SiO$_2$ systems are given in Figure~\ref{fig:interfaces}. The~crystal structures of Pt~\cite{wyckoff1963wg}, Ni~\cite{swanson1953standard}, Si, and~SiO$_2$~\cite{nieuwenkamp1935kristallstruktur} are used to build the interfaces. For~Si the bc8 form given by Kasper~et~al.~\cite{kasper1964crystal} is used, and~we used the cristobalite SiO$_2$ of Nieuwenkamp~et~al., these two where specifically chosen to closely match each-others crystal lattice. The~top and bottom sections ($9\times9\times6$ Pt,~$10\times 10\times6$ Ni,~$3\times3\times2$ Si and $4\times4\times2$ SiO$_2$), are attached to strongly coupled thermostats ($\tau$ = 100 fs) and act as a heat source and heat sink. The~intermediate regions (\mbox{$9\times9\times9$ Pt}, $10\times 10\times10$ Ni,~$3\times3\times25$ Si and $4\times4\times24$ SiO$_2$) are weakly coupled ($\tau$ = 10$^5$ fs). The~supercells are initially placed at a small distance and approach each other during an energy minimization to form the interface. From~these energy minimized (merged) systems, the~simulations are started. In~the non-reactive systems, the~top sections are kept at \mbox{$T_H$ = 330 K} and bottom sections are kept at \mbox{$T_C$ = 300 K}. In~the reactive systems temperature values are varied to trigger a chemical reaction at the interface. To~calculate the TBR, all the simulations are done over 1 ns, from~which the last 0.75 ns are considered to obtain the~data.
When an interface between two different materials is created artificial mechanical stresses are introduced by fitting the different lattice constants in one single periodic box. To~restrict this to a minimum we carefully selected the materials, supercells, and~orientation to create the interface. Thereby, the~deformation of the crystals is limited to 0.1\% compared to their literature value. Furthermore, we studied the influence of 1\% deformation of Platinum on the thermal conductivity. The~thermal conductivity for a compressed, as~well as, a~stretched crystal was lower, however for both cases within the standard deviation of the original system. The~deformation of the crystals, and~the study towards the thermal conductivity can be found in Appendix \ref{app:mech_defer}.%reference to appendix C
\section{Results}\label{sec:results}
ReaxFF MD simulations are carried out to understand the temperature discontinuity across the solid interfaces of homogeneous (Pt/Pt), heterogeneous (Pt/Ni) and heterogeneous reactive materials (Si/SiO$_2$). The~computed material properties from the previous section are used to upscale the results from the molecular level to macroscopic phenomenological-theory level. For~the thermal conductivity, the~extrapolated value is used. A~comparison is made between both methods for the non-reactive~interfaces. 
\subsection{Non-Reactive~Interfaces}
In realistic experimental conditions, the~thermostat takes time to set the desired temperature, thus~the temperature of the heat source evolves with time. To~study this effect on the final temperature profile in ReaxFF MD, we compared two different settings to increase the temperature of the heat source. One with a gradual temperature increase to $T_H$, and~one with an instantaneously high temperature at $T_H$. See Appendix~\ref{app:comp}, and Figure~\ref{fig:compTemp}, for the result of the comparison between the two cases. We observe that the final temperature profile is almost the same for both cases. Thus in the following cases, we have initialized the temperature of the heat source instantaneously at high temperature (instant $\Delta$$T$).

\begin{figure}[H]
	\centering
	\includegraphics[scale=0.95]{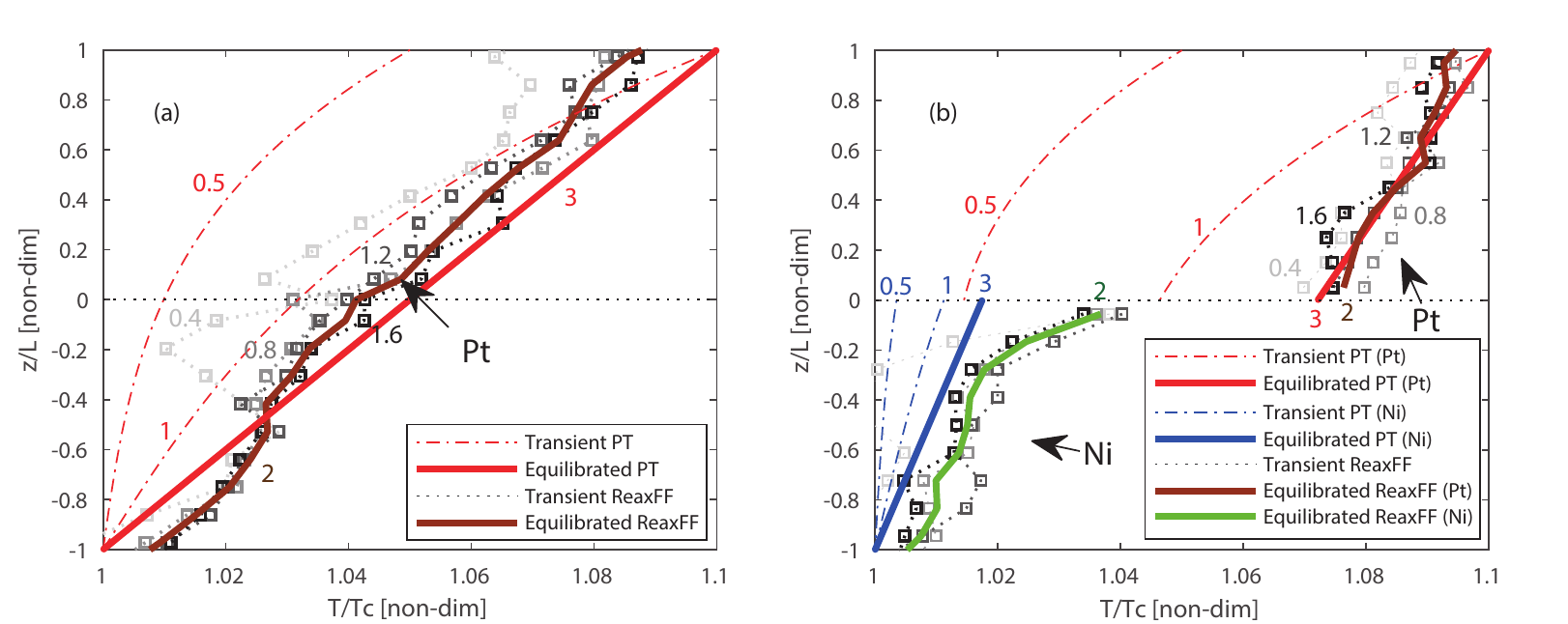}
	\caption{\label{fig:interfaces_PtPt_PtNi} Comparison of temperature profile computed from PT based model and ReaxFF MD simulation for: (\textbf{a}) Pt/Pt interface. (\textbf{b}) Pt/Ni interfaces. The~solid lines represent the equilibrated, and~the dashed/dotted lines the transient temperature profiles. The~dashed-dotted lines are intermediate transient temperature profiles for PT, and~the dotted lines with square markers represent intermediates from ReaxFF MD simulations. The~corresponding numbers at the lines represent $t_\text{non-dim}$. }
\end{figure}

For the non-reactive ReaxFF MD interface investigation, the~systems given in Figure~\ref{fig:interfaces}a,b are studied. The~temperature profile evolution across the solid interfaces of Pt/Pt and Pt/Ni systems with ReaxFF MD is plotted in gray-scale after every 200 ps, which can be seen in Figure~\ref{fig:interfaces_PtPt_PtNi}a,b. The light-gray to black lines represents respectively the earlier and later time~periods. 
The equilibrated temperature is represented by the solid brown and green lines, respectively Platinum and Nickel. The~temperature profiles developed over time obtained from ReaxFF MD are compared with the temperature profile from phenomenological-theory as shown in Figure~\ref{fig:interfaces_PtPt_PtNi}. The~molecular level ReaxFF MD simulations and macroscopic level phenomenological-theory have a different time scale, thus to compare them, a~non-dimensional time ($t_{\text{non-dim}}$) is defined as:
\begin{equation}\label{Tnon}
t_{\text{non-dim}} = t/t_{eq}
\end{equation}
where $t$ is the actual time and $t_{eq}$ is the time assumed the system is in steady state. An~steady state time of 0.5 ns is assumed for the MD simulations. The~red, blue and gray numbers in the figures represent these different transient $t_{\text{non-dim}}$ states. The~temperature profile obtained from the phenomenological-theory evolves slowly with time when compared with the ReaxFF MD simulations, where the transient states are quickly converging and fluctuation around the equilibrium state. In~the ReaxFF MD results, a~continuous temperature profile is observed at the Pt/Pt interface while a temperature jump (discontinuity) is observed at the Pt/Ni interface as shown in Figure~\ref{fig:interfaces_PtPt_PtNi}a,b, respectively. Similar temperature profiles are also observed from the phenomenological-theory. The~ReaxFF MD method shows a temperature jump of approximately 39\%, where the phenomenological-theory results in a 55\% jump of the imposed temperature difference of 30 K. Since one method is based on molecular level and another one is based on macroscopic level theory, a~slight discrepancy in the magnitude of the temperature jump can be expected. These results confirm that the temperature jump is observed at the solid interface between different materials for both molecular level and macroscopic level~modeling.

\subsection{Reactive~Interfaces}
At the surface of re-entry vehicles chemical reactions can occur, these reactions contribute to the heating of such  vehicles~\cite{giordano2017exploratory,kulkarni2012oxygen}. Furthermore, these reactions form a small layer, and~influence the heat and mass transport at the surface. To~gain more fundamental knowledge of such a surface, we~studied a reactive solid Si/SiO$_2$ interface (see Figure~\ref{fig:interfaces}c,d). The~building of the physical system is similar to the two previous systems (Pt/Pt and Pt/Ni). The~length between the heat source and sink is approximately 327 \AA, and~temperatures of the heat source and sink are varied to increase/decrease the chemical reaction at the solid interface~\cite{tromp1985high}. Four different cases are studied, for~the first case (Case~1) the set temperatures for heat source ($T_H$) and sink ($T_C$) are 150 and 100 K, respectively. For~the second case (Case~2), the~temperatures are $T_H=$ 350 K and $T_C$ = 300 K, and~for the third case (Case~3) the temperatures are $T_H$ = 850 K and $T_C$ = 800 K. These cases have an interface temperatures of approximately 125, 325, and~825 K. In~the fourth case (Case 4) the complete system was heated to a high temperature (1700 K) for 3.5 ns to create a reactive region, and~thereafter, cooled back to \mbox{$T_C$ = 100 K}, and~$T_H$ = 150 K to stop the chemical activity completely again. After~the cooling, a~new steady state simulation was done at $T_C$ = 100 K, and~$T_H$ = 150 K. {From Figure~\ref{fig:SiSiO2_interface} in Appendix~\ref{app:Si--SiO_interface}, one~can see that the major part of the interface formation takes place within the first nanosecond. In~terms the thickness of the interface, as~well as, the~depth of the oxygen penetration into the silicon surface only little changes are observed after the first nanosecond. Therefore, it was not needed to go for longer simulations to create the interface.} This fourth case was chosen to get a distinct comparison between the non-reacted and reacted interface, at~the same temperature (Case 1 and 4).

\begin{figure}[H]
	\centering
	\includegraphics[scale=0.95]{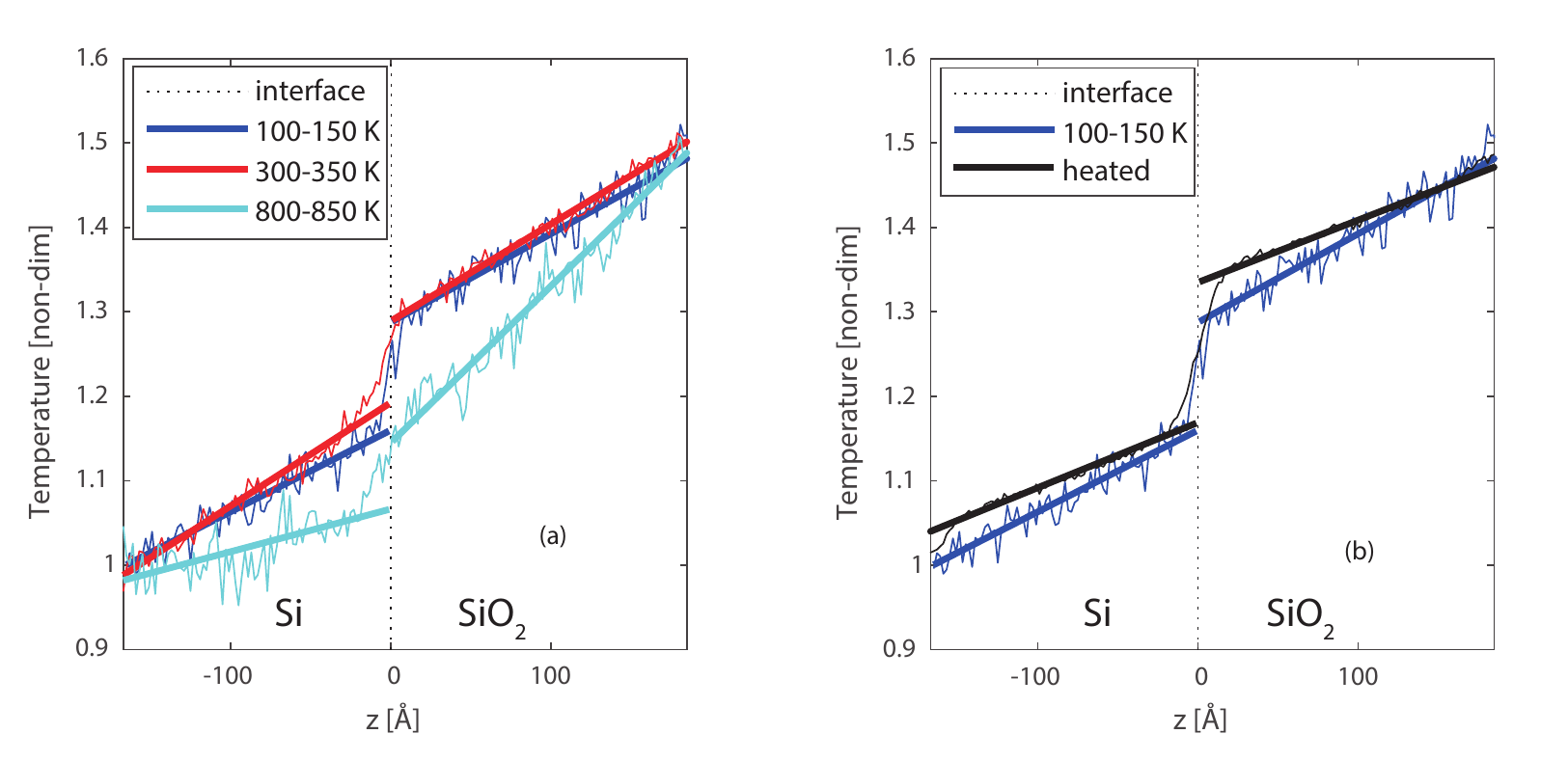}
	\caption{\label{fig:interfacesSiSiO2}Temperature profiles for the Si/SiO$_2$ interfaces, with~(\textbf{a}) case 1, case 2, and~case 3, and~(\textbf{b}) case~1, and~case 4. The~dotted line indicates the location of the interface, the~dark blue lines represent case 1 ($T_H$ = 150 and $T_C$ = 100 K), the~red lines represent case 2 ($T_H$ = 350 and $T_C$ = 300 K), the~light blue lines represent case 3($T_H$ = 850 and $T_C$ = 800 K), and~the black line represents case 4 (first heated to 1700 K, and~then a steady state at $T_H$ = 150 and $T_C$ = 100 K).}
\end{figure}

The resulted temperature profiles for Case 1--3 are plotted in Figure~\ref{fig:interfacesSiSiO2}a. The~temperature is made non-dimensional by taking 100--150 K as reference and divide by 100. The~thick solid lines are fitted to the data, and~extrapolated to the interface, to~get the temperature jump. The~initial interface is positioned at $z$ = 0 \AA~in the figures. A~temperature jump is observed between Si and SiO$_2$ for all the cases. It reiterates that there is a temperature discontinuity at the reactive heterogeneous solid interface as well. Case 1 (100--150 K) shows a clear jump, with~a sharp temperature profile. When the temperature is increased to 300--350 K (case 2), the~jump remains, however, it is less sharp. This is caused by some small deformation at the interface induced by the temperature. Moving to even higher temperatures 800--850 K (case 3), not only a temperature jump but also a drop of the temperature profile over the entire system is observed. This suggests a heat sink at the interface, due to energy consumption by the deformation of the crystals at the interface. This deformation has an impact on the heat transfer and results in an intermediate region between the Silicon and Silica crystals of a few~\aa{}ngstr\"{o}ms.

\begin{figure}[H]
	\centering
	\includegraphics[scale=0.9]{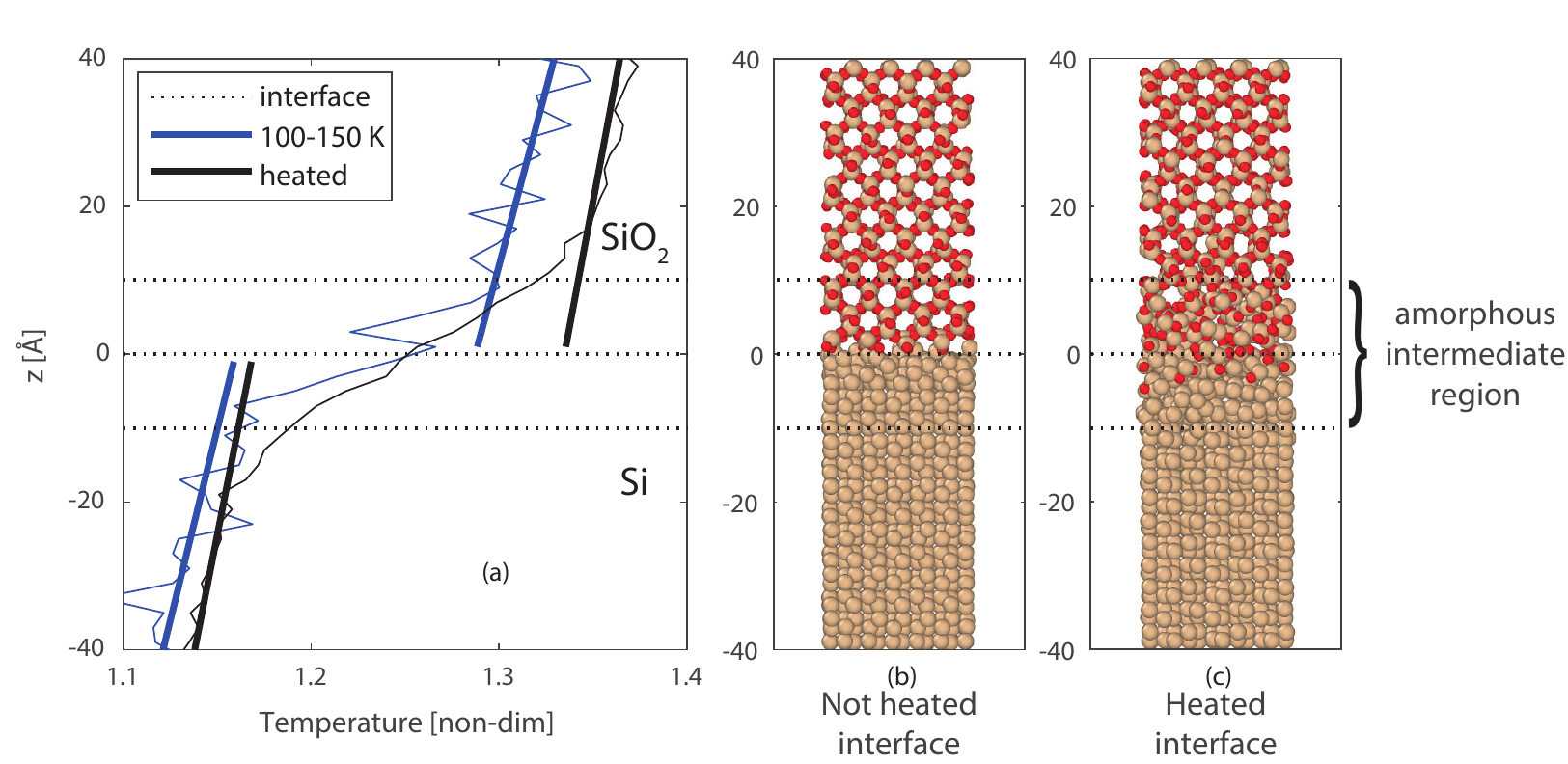}
	\caption{\label{fig:interfacesSiSiO2_large}(\textbf{a}) zoom of Figure~\ref{fig:interfacesSiSiO2}b, with~the temperature profiles for case 1 ($T_H$ = 150 and $T_C$ = 100 K), and~case 4 (first heated to 1700 K, and~then a steady state at $T_H$ = 150 and $T_C$ = 100 K). The~final molecular structures of the interfaces are given in (\textbf{b}) for case 1, and~(\textbf{c}) for case~4.}
\end{figure}

To better study the effect and size of the intermediate region, the~entire Si/SiO$_2$ system was heated up to 1700 K to increase the reactivity and advance the formation of the intermediate region. Higher temperatures were also tested, however, these resulted in melting of the Silicon crystal, and~separation of the two slabs. Lower temperatures would require more time to create a similar intermediate region. The~heating process results in a larger intermediate amorphous region, where oxygen diffused up to 5 \AA{} into the Silicon crystal, and~deformation of both materials is visible up to 10 \AA{} from the interface. After~the heating, the~system was cooled back to 100--150 K, this temperature was chosen to stop the chemical activity as far as possible. The~final equilibrated temperature profile is compared with the temperature profile of case 1, where no reactive region is had been present at the interface. This comparison shown in Figure~\ref{fig:interfacesSiSiO2}b, and~a closer profile around the interface is shown in Figure~\ref{fig:interfacesSiSiO2_large}a--c. From~Figure~\ref{fig:interfacesSiSiO2_large}c, the~thicker interfacial region for the reacted interface can be clearly observed, compared to the non-reacted clean interface (Figure~\ref{fig:interfacesSiSiO2_large}b). For~the heated interface (case 4), the~temperature jump ($\Delta T$) is larger, however, is has become less sharp than case 1 and smoothed over the formed intermediate~region.

The thermal boundary resistances (TBR) are given in Table~\ref{tab:Si_SiO2_gradients}. The~calculated value for the low temperature (case 1) is in good agreement with Deng~et~al.~\cite{deng2014kapitza}, who found a value of $1.48~(\pm0.46) \times 10^{-9}$ m$^2$K/W using NEMD, and~$1.37~(\pm0.42) \times 10^{-9}$ m$^2$K/W using phonon wave-package dynamics approach. 
The higher temperatures (case 2,3) approach the experimental results of Hurley~et~al. \cite{hurley2011measurement}, who measured a resistance of $2.3\times 10^{-9}$ m$^2$K/W. {The reacted interface, which includes an amorphous SiO$_2$ region, is in good agreement with the DMM results of Hu~et~al.\cite{hu2001thermal}, who found a resistance of $3.5\times 10^{-9}$ m$^2$K/W, for~amorphous SiO$_2$ with Si}. The~temperature jump at the interface for the 300--350~K and 800--850 K temperatures (case 2,3) is smaller, however, the~calculated TBR is higher, caused by the deformation of the interface which acts as an extra heat sink.
The TBR for the reacted interface (case 4) is more than twice the TBR of the non-reacted clean interface (case 1) at the same temperature, caused by the amorphous Si--SiO$_2$ intermediate~region.

\unskip
\begin{table}[H]
	\centering
	\caption{\label{tab:Si_SiO2_gradients}Thermal Boundary Resistance (TBR) of Si/SiO$_2$ systems, with~$T_C$ and $T_H$ (K) as thermostatted temperature for the heat sink and source respectively. }
	\begin{tabular}{ccccc}
	\toprule
		\textbf{Case}  & \textbf{\boldmath{$T_C$ (K)}} &  \textbf{\boldmath{$T_H$ (K)}} &  \textbf{\boldmath{Temperature Jump, $\Delta T$ ($K$)}}& \textbf{TBR}  \textbf{\boldmath{$\left(\dfrac{\text{m}^{2}\text{K}}{\text{W}}\right)$}}\\ \midrule   
		1  & 100 & 150  & 12.8  & $ 1.65\times 10^{-9}$  \\ 
        2  & 300 & 350  & 9.7   & $ 2.40\times 10^{-9}$  \\  
        3  & 800 & 850  & 7.8   & $ 2.34\times 10^{-9}$\\ 
		4  & 1700 $\rightarrow $ 100 & 1700 $\rightarrow $ 150 & 16.6  &$ 3.38\times 10^{-9}$\\ \bottomrule  
	\end{tabular} 
\end{table}

\section{Conclusions}\label{sec:conclusion}
ReaxFF MD is known to capture the physical and chemical phenomena under various conditions~\cite{senftle2016reaxff,pathak2016reactive}. We have chosen various force fields for Pt/Ni and Si/SiO$_2$ systems, which can mimic their material properties. The~selected force fields predict the equilibrium volumes~\cite{wyckoff1963wg,swanson1953standard,kasper1964crystal,nieuwenkamp1935kristallstruktur} and bulk modulus~\cite{holmes1989equation,chen2000compressibility,PhysRevBSiDFT,liu1993bulk} of respectively Pt, Ni, Si and SiO$_2$ in close agreement with experiments/theory. To~validate further, thermal expansion and thermal conductivity coefficients of Pt and Ni are estimated using ReaxFF MD. The~thermal expansion coefficients are found to be in reasonable agreement with experiments. The~thermal conductivity of a solid material is size-dependent on the molecular level. Thus, we have obtained the thermal conductivity of Pt and Ni for various system sizes and extrapolated to a very long length to determine the bulk thermal~conductivity. 

The elastic and thermal properties, obtained from ReaxFF MD, served as input parameters for the macroscopic level phenomenological-theory (PT)~\cite{giordano2017exploratory}. Temperature profiles of non-reactive interfaces, obtained from both methods, are compared. In~this comparison, we have reported the temperature profiles across a homogeneous (Pt/Pt), and~heterogeneous (Pt/Ni) interface. Temperature continuity is observed at the solid homogeneous interface of Pt/Pt. The~temperature profile of the molecular level simulation is faster at equilibrium than the phenomenological-theory. The~temperature profiles between Pt/Ni has a discontinuity at the interface observed in both molecular and macroscopic level. The~temperature jump obtained from the molecular level calculation is 18\% lower than the one obtained from PT calculations. The~discrepancy between the two models in the temperature jump for Pt/Ni is minimal, and~can be explained by the fact that the length- and time-scale for both calculations are different, and/or the length dependence of the thermal conductivity. We can conclude that both models, the~molecular level ReaxFF MD simulations and the PT, predict a temperature discontinuity across the solid boundary if the materials are not the~same. 

The ReaxFF MD methodology can capture chemical reactions, therefore, interesting insights could be obtained for solid pairs which can form a reactive interface. For~this purpose, the~Si/SiO$_2$ pair was chosen and the heat source and heat sink were varied to increase/decrease the chemical reaction at the interface. Three distinct solids (Si, amorphous reacted Si--SiO$_2$ interface, and~SiO$_2$) have been observed. The~thermal boundary resistance (TBR) is computed at the Si/SiO$_2$ interface for the different systems, providing us with information of the TBR over interfaces with different chemical activity. It can be concluded that the reacted amorphous region at the interface introduces extra resistivity, compared to the non-reactive clean interface. Showing the opportunity to control the thermal resistivity of a multi-layered system by controlling the interfacial reactive~regions.

%%%%%%%%%%%%%%%%%%%%%%%%%%%%%%%%%%%%%%%%%%
\vspace{6pt} 

%%%%%%%%%%%%%%%%%%%%%%%%%%%%%%%%%%%%%%%%%%
%% optional
%\supplementary{The following are available online at \linksupplementary{s1}, Figure S1: title, Table S1: title, Video S1: title.}


%%%%%%%%%%%%%%%%%%%%%%%%%%%%%%%%%%%%%%%%%%
\authorcontributions{conceptualization, S.N. and D.G.; methodology, S.N. and D.G.; validation, K.H., A.P. and P.S.; investigation, K.H., A.P. and P.S.; resources, D.S.; writing--original draft preparation, K.H., A.P. and S.N.; writing--review and editing,  K.H., A.P., S.N., P.S., D.G. and D.S.; supervision, S.N., D.G. and D.S.; funding acquisition, D.S.}

%%%%%%%%%%%%%%%%%%%%%%%%%%%%%%%%%%%%%%%%%%
\funding{This research received no external funding.}

%%%%%%%%%%%%%%%%%%%%%%%%%%%%%%%%%%%%%%%%%%
\acknowledgments{The authors would like to thank the Netherlands Organization for Scientific Research (NWO) for access to the national high performance computing facilities (Dossiernr: 17092 6026).}

%%%%%%%%%%%%%%%%%%%%%%%%%%%%%%%%%%%%%%%%%%
\conflictsofinterest{The authors declare no conflict of~interest.} 

%%%%%%%%%%%%%%%%%%%%%%%%%%%%%%%%%%%%%%%%%%
%% optional
%\abbreviations{The following abbreviations are used in this manuscript:\\
%
%\noindent 
%\begin{tabular}{@{}ll}
%MDPI & Multidisciplinary Digital Publishing Institute\\
%DOAJ & Directory of open access journals\\
%TLA & Three letter acronym\\
%LD & linear dichroism
%\end{tabular}}

%%%%%%%%%%%%%%%%%%%%%%%%%%%%%%%%%%%%%%%%%%
%% optional
\appendixtitles{yes} %Leave argument "no" if all appendix headings stay EMPTY (then no dot is printed after "Appendix A"). If~the appendix sections contain a heading then change the argument to "yes".
\appendix
\section{Birch-Murnaghan Equation of State Fitting Si \& SiO$_2$}\label{app:BM-eos}
To select the correct force field, a~BM-EOS study is done. The~results of the fitting are presented in Table~\ref{tab:app_BM}, with~$r^2$ as correlation coefficient of the fitting. The~results are compared with literature, which gives $V_{0,\text{Pt}}= 60.38~\text{\AA}^3$~\cite{wyckoff1963wg}, $V_{0,\text{Ni}}= 43.76~\text{\AA}^3$~\cite{swanson1953standard}, $V_{0,\text{Si}}= 292~\text{\AA}^3$~\cite{kasper1964crystal}, $V_{0,\text{SiO}_2}= 171~\text{\AA}^3$~\cite{nieuwenkamp1935kristallstruktur}, $B_{0,\text{Pt}}= 266$ GPa~\cite{holmes1989equation}, $B_{0,\text{Ni}}=185$ GPa~\cite{chen2000compressibility}, $B_{0,\text{Si}}=98$ GPa~\cite{hopcroft2010young}, and~ $B_{0,\text{SiO}_2}=36$ GPa~\cite{liu1993bulk}. The~ReaxFF of Mueller~et~al.~~\cite{mueller2010development} and Kulkarni~et~al.~~\cite{kulkarni2012oxygen} proves to be the best applicable for respectively the Pt/Ni system and the Si/SiO$_2$ system.
\begin{table}[H]
	\centering
		\caption{\label{tab:app_BM}Results of BM-eos fitting to different~ReaxFF.}
		\begin{tabular}{ccccccc}
		\toprule
			& \textbf{Reference} & \textbf{\boldmath{$B_0$}} & \textbf{\boldmath{$V_0$}} & \textbf{\boldmath{$E_0$}} & \multirow{2}{*}{\textbf{\boldmath{$ r^2 $}}} &\multirow{2}{*}{\textbf{Figure}}\\ 
			& \textbf{Force Field} &  \textbf{(GPa)}   & \textbf{\boldmath{$ ($\AA$^3)$}}&  \textbf{(kcal/mol)} &  & \\
			\midrule Pt
			& \cite{mueller2010development} & 240  & 61.52 & $-$532.4 & 1.0 & Figure \ref{fig:BMeos}a L1\\ 
			& \cite{sanz2008molecular} & 179  & 62.91 & $-$534.3 &  1.0 &Figure~\ref{fig:BMeos}a L2\\ 
			& \cite{nielson2005development} & 166  & 64.52 & $-$560.6 &  1.0  &Figure~\ref{fig:BMeos}a L3\\ 
		    \midrule Ni 
			& \cite{mueller2010development} & 155  & 46.96 & $-$414.9 &  1.0 &Figure~\ref{fig:BMeos}b L2\\ 
			& \cite{nielson2005development} & 167  & 48.21 & $-$369.8 &  1.0  &Figure~\ref{fig:BMeos}b L1\\ 
			\midrule Si 
			& \cite{kulkarni2012oxygen} &  144 & 272.1 & $-$1617 & 0.92 & Figure~\ref{fig:BMeos}c\\ 
			& \cite{fogarty2010reactive} &  165 & 273.6 & $-$1611 & 0.88  &\\ 
			& \cite{narayanan2011reactive} &  292 & 252.9 & $-$1729 & 0.95 &\\ 
			& \cite{newsome2012oxidation} &  216 & 291.2 & $-$1675 & 0.53 & \\ 
			& \cite{nielson2005development} &  295 & 265.7 &$-$2244 & 0.84  &\\ 
			& \cite{rahnamoun2014reactive} &  235 & 268.3 & $-$2206 & 0.95  &\\ 
			& \cite{zou2012investigation} &  334 & 255.2 & $-$1728 & 0.95 &\\ 
			& \cite{pitman2012dynamics} &  289 & 252.3 & $-$1730 & 0.95 & \\ 
			& \cite{castro2013comparison} &  334 & 255 & $-$1728 & 0.95  &\\ 
			\midrule SiO$_2$  
			& \cite{kulkarni2012oxygen} &  35  & 242.1 & $-$2128 & 0.83 &Figure~\ref{fig:BMeos}d\\ 
			& \cite{fogarty2010reactive} &  500 & 244.6 & $-$1847 & 0.60 & \\ 
			& \cite{narayanan2011reactive} &  33  & 260.0 & $-$1818.6 & 0.86  &\\ 
			& \cite{newsome2012oxidation} &  34  & 238.1& $-$1793 & 0.98 &\\ 
			& \cite{nielson2005development} &  47  & 244.7 & $-$1860 & 0.83  &\\ 
			& \cite{rahnamoun2014reactive} &  234 & 247.2 & $-$1828 & 0.78  &\\ 
			& \cite{zou2012investigation} &  331 & 263.4 & $-$1841& 0.19 &\\
			& \cite{pitman2012dynamics}  &  273 & 258.1 & $-$1837 & 0.16 &\\  
			& \cite{castro2013comparison} &  331 & 263.4 &$-$1840 & 0.19 &\\ \bottomrule
		\end{tabular} 
\end{table}

\section{Finite Size Effects, Perpendicular to the Heat~Flow}\label{app:size-effects_perp}
Platinum systems ($3\times 3\times 32 $, $5\times 5\times 32 $, $10\times 10\times 32 $) were used to study the finite size effects in perpendicular direction to the heat flow. The~system are placed in vacuum in z-direction and have periodic boundary conditions in x- and y-direction, the~given crystal sizes are including heat source and sink. No final size effects are observed in perpendicular direction to the heat~flow.
\begin{table}[H]
	\centering
	\caption{\label{tab:Pt_perp_size}Thermal conductivity of Pt-systems in vacuum in z-direction, and~different sizes in x- and~y-direction.}
	\begin{tabular}{cc}
		\toprule 
		\textbf{System} & \textbf{Thermal Conductivity (W/mK)}\\ \midrule
		$3\times 3\times 32 $& 9.7 \\
		$5\times 5\times 32 $& 8.0 \\
		$10\times 10\times 32$ & 10.7 \\ \bottomrule
	\end{tabular} 
\end{table}
\unskip

\section{Influence of Mechanical Deformation of~Slabs}\label{app:mech_defer}
When a heterogeneous interface with two different materials, and~thus different lattice parameters, is created the materials are compressed or stretched to fit both materials within the same periodic box. This introduces extra mechanical stresses in the crystals. To~restrict this to a minimum we have chosen the materials and supercells in such a way that these artificial deformations are kept to a minimum. The~lattice parameters are given in Tables~\ref{tab:PtNi} and~\ref{tab:SiSiO2} for the literature value, after~an energy minimization in ReaxFF, after~an energy minimization in ReaxFF with a vacuum in z direction, and~the size used in this work. The~lattice parameters are compared with the literature value and the error is given in the last column, this shows that the values are all within 3\%, and~there is only 0.1\% deformation in this work compared to the lattice from~literature.

\begin{table}[H]
	\centering
	\caption{\label{tab:PtNi}Lattice constant for Pt and~Ni.}
			\begin{tabular}{ccc}	
			\toprule
			\textbf{System} & \textbf{Lattice [a; b; c] (\AA)}& \textbf{Deviation from Literature (\%)}\\ \midrule
			Pt - literature~\cite{wyckoff1963wg} & 3.9231; 3.9231; 3.9231 & - \\ %Please check if this should be a minus sign or a em dash.
			Pt - EM ReaxFF &3.9473; 3.9473; 3.9473  & +0.6 \\
			Pt - EM ReaxFF + vacuum in z-direction & 3.9412 ; 3.9412 ; --- & +0.5 \\			
			Pt - interface & 3.9192; 3.9192; ---  & $-$0.1 \\ \midrule
			Ni - literature~\cite{swanson1953standard} & 3.5238; 3.5238; 3.5238 & --- \\
			Ni - EM ReaxFF & 3.6122; 3.6122; 3.6122 &  +2.5\\
			Ni - EM ReaxFF + vacuum in z-direction & 3.6048 ; 3.6048 ; --- & +2.3 \\			
			Ni - interface & 3.5273; 3.5273; ---  & +0.1 \\
			\bottomrule
		\end{tabular} 
	
\end{table}
\unskip

\begin{table}[H]
	\centering
	\caption{\label{tab:SiSiO2}Lattice constant for Si (bc8), and~cristobalite SiO$_2$.}
		\begin{tabular}{ccc}
		\toprule
			\textbf{System} & \textbf{Lattice [a; b; c] (\AA)} & \textbf{Deviation From Literature (\%)}\\ \midrule
			Si (bc8) - literature~\cite{kasper1964crystal} & 6.636; 6.636; 6.636 & --- \\
			Si (bc8) - EM ReaxFF & 6.4393; 6.4393; 6.4393  & $-$2.9 \\
			Si (bc8) - EM ReaxFF + vacuum in z-direction & 6.4383; 6.4046; --- & $-$2.9; $-$43.5 \\			
			Si (bc8) - interface & 6.6273; 6.6273; --- & $-$0.1 \\ \midrule
			SiO$_2$ - literature~\cite{nieuwenkamp1935kristallstruktur} & 4.964; 4.964; 6.920 & --- \\
			SiO$_2$ - EM ReaxFF & 5.0443; 5.0443; 7.0063 & +1.0 \\
			SiO$_2$ - EM ReaxFF + vacuum in z-direction & 5.0443; 5.0443; --- & +1.0 \\			
			SiO$_2$ - interface & 4.9705; 4.9705; --- & +0.1 \\
		\bottomrule
		\end{tabular} 
\end{table}

\subsection*{Influence of Stress on Thermal~Conductivity}
To form an interface with different materials, the~materials are slightly stressed to match each other lattice constants. One of the two materials was slightly compressed, and~the other one slightly stretched to form the interface. In~the Tables~\ref{tab:PtNi} and \ref{tab:SiSiO2} amount of deformation is shown for the materials used in this work, which are within $\pm 0.1\%$. To~gain more knowledge on the effect of these stresses on the heat transport across the material, we computed the thermal conductivity for \mbox{3 $\times$ 3 $\times$ 132 Platinum} structures with lattices corresponding to the literature value, 1\% compressed structures, a~1\% stretched structures. The~computed thermal conductivities for the different deformation are given in Table~\ref{tab:Pt_stress}, and~are within each other’s standard deviation. Thereby, we conclude that we can neglect the effect of stress in this~study.
\begin{table}[H]
	\centering
	\caption{\label{tab:Pt_stress}Thermal conductivity of a 3  $\times$  3  $\times$  132 Platinum structure under different mechanically induced~stresses.}
		\begin{tabular}{cccc}
		\toprule
			  \multirow{2}{*}{\textbf{Simulation} }       & \textbf{No Stress} & \textbf{1\% Compression} & \textbf{1\% Stretched}\\ 
			 & \textbf{k (W/mK)}& \textbf{k (W/mK)} &\textbf{ k (W/mK)}\\ \midrule
			1 & 15.2 & 14.6 & 13.4\\
			2 & 17.8 & 13.9 & 14.6\\
			3 & 16.7 & 15.3 & 17.4\\
			4 & 15.6 & 16.5 & 16.5\\ \midrule
			Average & 16.3 $\pm$ 1.2 & 15.1 $\pm$ 1.1 & 15.5 $\pm$ 1.8 \\\bottomrule
		\end{tabular} 
\end{table}

\section{Comparison of  Gradual and Instant Induced~Temperature}\label{app:comp}
\begin{figure}[H]
	\centering
	\includegraphics{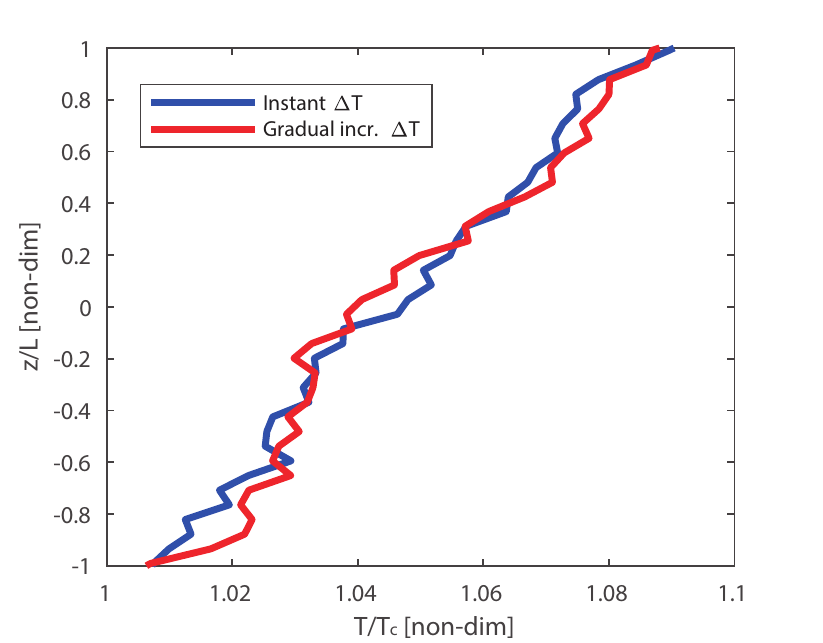} 
	\caption{\label{fig:compTemp}Comparison of temperature profiles (non-dim. \emph{T} versus non-dim. \emph{x}) obtained from ReaxFF-MD simulations for the Pt/Pt interface in the case where the heat source is immediately heated to $T_H$ = 330 K (instant $\Delta$$T$) and system in which the heat source is gradually heated to $T_H$ = 330 K in steps of 5 K per 1 $\times$ 10$^5$ iterations (gradual $\Delta$$T$).}
\end{figure}

In realistic experimental conditions, the~thermostat takes time to set the desired temperature, thus the temperature of the heat source evolves with time. To~investigate the effect of heating the systems on the final temperature distribution in ReaxFF-MD, we have compared a gradual temperature increase to $T_H$ and an instantaneously temperature at $T_H$. For~the gradual temperature setting, we have increased the temperature of the heat source ($T_H$ = 330 K) in the steps of 5 K per 0.1 million iterations (25 ps). After~0.6 million iterations (150 ps), the~temperature profile of the gradual temperature rise system is compared with the system in which temperature of hot zone was instantaneously set at \mbox{$T_H$ = 330 K} (instant $\Delta$$T$) as shown in Figure~\ref{fig:compTemp}, the~comparison was done over a total range of 1 million iterations. We observe that the equilibrated temperature profile is almost the same for both cases. Thus~in the following cases, we have initialized the temperature of the heat source instantaneously at high temperature (instant $\Delta$$T$). 
\unskip
\section{Development of the Amorphous Si/SiO$_2$ Interface}\label{app:Si--SiO_interface}
\begin{figure}[H]
	\centering
	\includegraphics[scale=0.95]{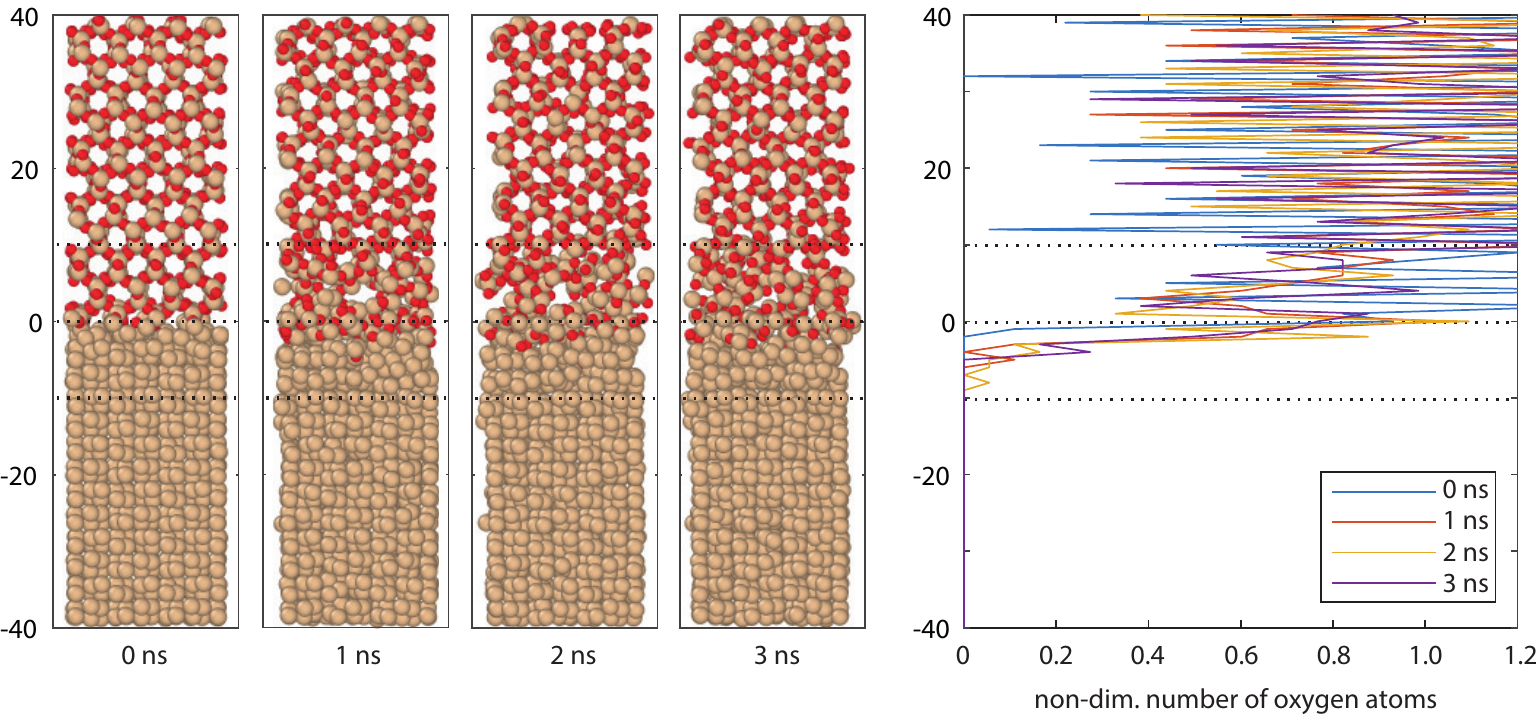} 
	\caption{\label{fig:SiSiO2_interface}Snapshots of the Si/SiO$_2$ interface.}
\end{figure}

To create a reactive formed interface between Si and SiO$_2$, the~system was heated for to high temperatures (1700 K) for 3.5 ns. Thereafter, it was cooled back to $T_C$ = 100 K, and~$T_H$ = 150 K to stop the chemical activity completely again. The~development of the interface over the first 3 ns can be observed in the snapshots in Figure~\ref{fig:SiSiO2_interface}, also the number of oxygen atoms are counted for these~snapshots.

%=====================================
% References, variant A: internal bibliography
%=====================================
%\reftitle{References}
%\begin{thebibliography}{999}
%% Reference 1
%\bibitem[Author1(year)]{ref-journal}
%Author1, T. The title of the cited article. {\em Journal Abbreviation} {\bf 2008}, {\em 10}, 142--149.
%% Reference 2
%\bibitem[Author2(year)]{ref-book}
%Author2, L. The title of the cited contribution. In {\em The Book Title}; Editor1, F., Editor2, A., Eds.; Publishing House: City, Country, 2007; pp. 32--58.
%\end{thebibliography}

% The following MDPI journals use author-date citation: Arts, Econometrics, Economies, Genealogy, Humanities, IJFS, JRFM, Laws, Religions, Risks, Social Sciences. For those journals, please follow the formatting guidelines on http://www.mdpi.com/authors/references
% To cite two works by the same author: \citeauthor{ref-journal-1a} (\citeyear{ref-journal-1a}, \citeyear{ref-journal-1b}). This produces: Whittaker (1967, 1975)
% To cite two works by the same author with specific pages: \citeauthor{ref-journal-3a} (\citeyear{ref-journal-3a}, p. 328; \citeyear{ref-journal-3b}, p.475). This produces: Wong (1999, p. 328; 2000, p. 475)

%=====================================
% References, variant B: external bibliography
%=====================================
%\externalbibliography{yes}
%\bibliography{reference}
\reftitle{References}

%%%%%%%%%%%%%%%%%%%%%%%%%%%%%%%%%%%%%%%%%%
%% optional
%\sampleavailability{Samples of the compounds ...... are available from the authors.}

%% for journal Sci
%\reviewreports{\\
%Reviewer 1 comments and authors’ response\\
%Reviewer 2 comments and authors’ response\\
%Reviewer 3 comments and authors’ response
%}

%%%%%%%%%%%%%%%%%%%%%%%%%%%%%%%%%%%%%%%%%%
\end{document}